\documentclass{emulateapj}
\newcommand{\target}{XTE\,J1118+480}
\newcommand{\RXTE}{\textit{RXTE}}

\slugcomment{Accepted for publication in the Astrophysical Journal}

\shorttitle{Variability in XTE J1118+480 in Outburst}
\shortauthors{Hynes et al.}

\begin{document}

%% LaTeX will automatically break titles if they run longer than
%% one line. However, you may use \\ to force a line break if
%% you desire.

\title{Further Evidence for Variable Synchrotron Emission in XTE
J1118+480 in Outburst}

\author{R. I. Hynes\altaffilmark{1,7}, E. L. Robinson\altaffilmark{2},
K. J. Pearson\altaffilmark{1,7}, D. M. Gelino\altaffilmark{3,7}, 
      W. Cui\altaffilmark{4}, Y. Xue\altaffilmark{4}, 
      M. A. Wood\altaffilmark{5},  
      T. K. Watson\altaffilmark{6}, D. E. Winget\altaffilmark{2},
            I. M. Silver\altaffilmark{5}}
\altaffiltext{1}{Department of Physics and Astronomy, Louisiana State
University, Baton Rouge, Louisiana 70803, USA; rih@phys.lsu.edu}
\altaffiltext{2}{McDonald Observatory and Department of Astronomy, The
University of Texas at Austin, 1 University Station C1400, Austin,
Texas 78712} 
\altaffiltext{3}{Michelson Science Center, California Institute of
  Technology, MS100-22, Pasadena, California 91125} 
\altaffiltext{4}{Department of Physics, Purdue University, 525 Western
Avenue, West Lafayette, Indiana 47907}
\altaffiltext{5}{Department of Physics \& Space Sciences and SARA Observatory, Florida
Institute of Technology, Melbourne, FL 32901}
\altaffiltext{6}{Southwestern University, 1001 E. University Avenue,
Georgetown, TX 78626}
\altaffiltext{7}{Visiting
Astronomer, Kitt Peak National Observatory, National Optical Astronomy
Observatory, which is operated by the Association of Universities for
Research in Astronomy, Inc. (AURA) under cooperative agreement with
the National Science Foundation.}

\begin{abstract}

We present simultaneous multicolor infrared and optical photometry of
the black hole X-ray transient \target\ during its short 2005 January
outburst, supported by simultaneous X-ray observations.  The
variability is dominated by short timescales, $\sim10$\,s, although a
weak superhump also appears to be present in the optical.  The optical
rapid variations, at least, are well correlated with those in X-rays.
Infrared $JHK_{\rm s}$ photometry, as in the previous outburst,
exhibits especially large amplitude variability.  The spectral energy
distribution (SED) of the variable infrared component can be fitted
with a power-law of slope $\alpha=-0.78\pm0.07$ where $F_{\nu} \propto
\nu^{\alpha}$.  There is no compelling evidence for evolution in the
slope over five nights, during which time the source brightness
decayed along almost the same track as seen in variations within the
nights.  We conclude that both short-term variability, and longer
timescale fading, are dominated by a single component of constant
spectral shape.  We cannot fit the SED of the IR variability with a
credible thermal component, either optically thick or thin.  This IR
SED is, however, approximately consistent with optically thin
synchrotron emission from a jet.  These observations therefore provide
indirect evidence to support jet-dominated models for \target\ and
also provide a direct measurement of the slope of the optically thin
emission which is impossible based on the average spectral energy
distribution alone.

\end{abstract}


\section{Introduction}

The black hole X-ray transient (BHXRT) \target\ has proven to be a
crucial object for understanding the class.  It is one of the best
objects for multiwavelength study as it suffers an exceptionally low
interstellar extinction, allowing broad-band observations encompassing
even the extreme-UV region (\citealt{Hynes:2000a};
\citealt{McClintock:2001a}; \citealt{Chaty:2003a}).  It also stands
out in possessing the shortest period yet known among objects with
black hole primaries (4.1\,hr) and has exhibited only low luminosity
hard state outbursts, reaching just $1.2\times10^{36}$\,erg\,s$^{-1}$
\citep{McClintock:2001a}.

The increasingly dominant paradigm for understanding the emission from
\target, and from hard state BHXRTs in general, is that it (usually)
involves an evaporated hot inner disk launching a compact jet
(\citealt{Markoff:2001a}; \citealt{Yuan:2005a}).  The jet is
responsible for not only radio emission, but also much of the IR,
optical, and possibly even some UV emission.  The properties of the
unusual UV, optical, and IR (UVOIR) variability also point to
synchrotron emission rather than reprocessing of X-rays by a disk
(\citealt{Kanbach:2001a}; \citealt{Hynes:2003a}).  In fact it has even
been proposed that the X-ray emission seen in the hard state could
arise from synchrotron emission \citep{Markoff:2001a}, although this
is not widely accepted and may be overwhelmed by a larger Comptonized
X-ray component \citep{Yuan:2005a}.

Part of the difficulty in bridging from the UVOIR region to X-rays is
that the former is likely a mixture of jet and disk emission, with the
disk emission masking the break from flat-spectrum to optically thin
synchrotron.  Consequently the position of the break, and slope of the
optically thin component are not directly observable, increasing the
uncertainty in extrapolating to X-rays.  The variability may provide
the key to disentangling these components.  To test this possibility,
we have assembled serendipitous multiwavelength observations of
\target\ during the 2005 January outburst.  The 2005 outburst was a
much shorter and somewhat weaker event than that seen in 2000.  It was
discovered optically by \citet{Zurita:2005a}, with the first high
points seen on 2005 January 9.  The outburst was also detected at
X-ray and radio wavelengths (\citealt{Remillard:2005a};
\citealt{Pooley:2005a}; \citealt{Rupen:2005a}).  The outburst faded
rapidly, reaching near quiescence by late February
\citep{Zurita:2005b}.  For a discussion of the outburst properties and
longterm lightcurves see \citet{Zurita:2006a}.

\section{Observations}

\subsection{SARA 0.9\,m Optical Observations}

\target\ was observed using the SARA 0.9-m telescope located at
Kitt Peak National Observatory.  Observations used an Apogee AP7p CCD
camera with an $R$ filter.  Exposure times were 10\,s, with
approximately 7\,s intervening dead-time.  The raw data frames were
bias-, dark-, and flat field-corrected using standard {\sc iraf}
routines.  Once the data frames were calibrated, we extracted time
series aperture photometry using the external {\sc iraf} package
CCD\_HSP written by Antonio Kanaan (U.  Federal Santa Catarina,
Brazil).  CCD\_HSP automates the field alignment and photometry
extraction for time-series CCD data.

\begin{table*}
\small
\caption{Log of optical/IR observations}
\label{BigLogTable}
\begin{tabular}{lcc}
Telescope \& instrument & UT date \& time & Exposure (s) \\
\noalign{\smallskip}
\hline
\noalign{\smallskip}
SARA 0.9\,m, AP7p & 2005 Jan 13  06:03:04--13:50:50 & 10  \\
SARA 0.9\,m, AP7p & 2005 Jan 15  06:02:57--13:53:07 & 10  \\
\noalign{\smallskip}
McDonald 2.1\,m, Argos & 2005 Jan 13  07:42:07--09:04:39 & 1  \\
\noalign{\smallskip}
McDonald 2.7\,m, White Guider & 2005 Jan 16  07:51:06--12:28:23 & 3  \\
McDonald 2.7\,m, White Guider & 2005 Jan 19  09:17:12--10:56:48 & 3  \\
\noalign{\smallskip}
KPNO 2.1\,m, SQIID & 2005 Jan 15  09:59:49--14:00:29 & 2  \\
KPNO 2.1\,m, SQIID & 2005 Jan 16  08:01:48--13:52:09 & 2  \\
KPNO 2.1\,m, SQIID & 2005 Jan 17  10:28:01--13:51:59 & 1  \\
KPNO 2.1\,m, SQIID & 2005 Jan 18  07:54:30--13:54:12 & 2  \\
KPNO 2.1\,m, SQIID & 2005 Jan 19  09:39:40--13:53:08 & 2  \\
\end{tabular}
\end{table*}

\subsection{McDonald Observatory 2.1\,m Optical Observations}

\target\ was also observed using the Argos fast CCD
camera \citep{Nather:2004a} on the McDonald Observatory 2.1\,m
telescope.  About 1.5\,hrs of data were obtained 
in unfiltered white light.  The data were taken as a
continuous sequence of 1\,s images with negligible intervening
dead-time.  Conditions were mostly non-photometric with 1--2\,arcsec
seeing, so differential photometry was performed relative to the same
comparison star used in the IR.  Data reduction employed a combination
of {\sc iraf} routines to generate calibration files and then a custom
IDL pipeline to apply calibrations and extract photometry.  Bias
structure and dark current were subtracted using many dark exposures
of the same duration as the object frames.  Residual time-dependent
bias variations were removed using two partial bad columns which are
not light sensitive.  Unfortunately no flat fields were obtained in
the unfiltered mode, so no sensitivity corrections were applied.
Photometry was extracted using standard aperture photometry
techniques.

\subsection{McDonald Observatory 2.7\,m Optical Observations}

Additional optical observations of \target\ were performed 
using the White Guider CCD camera on the McDonald Observatory
2.7\,m telescope.  The data were taken as a continuous sequence
of 3\,s $R$ band images with about 2\,s of intervening dead-time.  Conditions
appeared near-photometric with 1--2\,arcsec seeing, so absolute
photometry was performed to maximize the signal-to-noise ratio.
Conditions deteriorated towards the end of the January 19 run, so
these data were discarded.  This approach was necessary as the
comparison star used with Argos was not available with this
instrument, and the brightest other comparison in the field was nearly
six magnitudes fainter than \target\ in outburst.  Initial data
reduction used standard {\sc iraf} procedures to remove bias and
flat-field the data.  Aperture photometry of \target\ used a 5$''$
aperture to minimize aperture losses.

\subsection{Kitt Peak National Observatory 2.1\,m Infrared Observations}
\label{IRObsSection}

Infrared photometry of \target\ was obtained 
using the Simultaneous Quad Infrared Imaging Device
(SQIID) on the 2.1\,m telescope at the Kitt Peak National Observatory.
Details are given in Table~\ref{BigLogTable} \& \ref{IRLogTable}.  The
camera records $J$, $H$, $K_{\rm s}$, and $L$ images simultaneously,
although we found that the $L$ data were of insufficient quality to be
useful.  $JHK_{\rm s}$ data were all of good quality, however, and
obtained simultaneously with 1\,s or 2\,s exposures separated by $\sim
54$\,s deadtime.  The image was nodded back and forth on alternating
exposures to facilitate sky subtraction.  The seeing was typically
better than 1.3\,arcsec in $J$.  Data reduction employed a combination
of {\sc iraf}\footnote{IRAF is distributed by the National Optical
Astronomy Observatories, which are operated by the Association of
Universities for Research in Astronomy, Inc., under cooperative
agreement with the National Science Foundation.} routines to generate
calibration files and then a custom IDL pipeline to apply calibrations
and extract photometry.

\begin{table*}
\small
\caption{Details of IR photometry}
\label{IRLogTable}
\begin{tabular}{lccccc}
UT date & \multicolumn{2}{c}{Number of images} & \multicolumn{3}{c}{Average magnitude of XTE~J1118+480} \\
                       & All & Good & $J$ & $H$ & $K_{\rm s}$ \\
\noalign{\smallskip}
\hline
\noalign{\smallskip}
2005 Jan 15  & 221 & 205 & 12.92 & 12.49 & 11.97 \\
2005 Jan 16  & 360 & 246 & 12.96 & 12.52 & 12.05 \\
2005 Jan 17  & 210 & 147 & 13.03 & 12.64 & 12.10 \\
2005 Jan 18  & 380 & 304 & 13.06 & 12.64 & 12.14 \\
2005 Jan 19  & 270 & 257 & 13.10 & 12.70 & 12.19 \\
\end{tabular}
\end{table*}

Where possible, sky subtraction was performed by subtracting the
average of immediately preceding and immediately following images.
Sensitivity variations were corrected using an average of many sky
flat images taken at twilight.  We verified that these images
acceptably flattened the sky background of target images before sky
subtraction.  Count rates were low enough that non-linearity was below
1\,\%, so no correction was applied.

Because of the source brightness during outburst, only one usable
comparison star was present in the field, and this was fainter than
our target.  We therefore used the comparison star only as a check of
the photometric stability and did not perform differential photometry.
Fortunately, conditions were mostly photometric, at least sufficiently
so that transparency variations are much smaller than the intrinsic
variability of the target.  The major exception occured at the end of
the third night, so these points were not used.

Our absolute calibration is based on 2MASS photometry of the
comparison star,\\2MASS~J11180724+4803527 \citep{Cutri:2003a}.  This
star has $J=13.449\pm0.024$, $H=12.825\pm0.028$, and $K_{\rm
s}=12.610\pm0.022$, where the uncertainties are dominated by
statistical effects and so should be uncorrelated.  We checked this
calibration against ARNICA standards AS19-0 and AS19-2
\citep{Hunt:1998a} after transforming the latter in the 2MASS
$JHK_{\rm s}$ system \citep{Carpenter:2001a}.  The standard yielded a
consistent calibration, with comparable uncertainties, so we retained
the 2MASS calibration for this work.  Our average calibrated
magnitudes for the target for each night are summarized in
Table~\ref{IRLogTable}.  The calibrated magnitudes were converted to
fluxes according to \citet{Cutri:2003a}.  After folding in
uncertainties in conversion from magnitudes to fluxes, we estimate
that the systematic uncertainties in our calibrated fluxes are then
2.8\,\%\ in $J$, 3.2\,\%\ in $H$ and 2.8\,\%\ in $K_{\rm s}$.

\subsection{Rossi X-ray Timing Explorer X-ray Observations}

\target\ was intensively monitored in X-rays with the {\em Rossi X-ray
  Timing Explorer} (\RXTE) from 2005 January 13 to February 26. There
  are two narrow-field instruments aboard \RXTE, the Proportional
  Counter Array (PCA) and the High-Energy X-ray Timing Experiment
  (HEXTE). Due to the faintness of the source throughout the outburst,
  only the more sensitive PCA produced useful data. For this work, we
  only used data taken in the {\em Standard1b} mode, which provides a
  time resolution of 1/8 sec and covers a nominal energy range of
  2--60 keV. The PCA data were reduced with {\em FTOOLS} (version 5.2)
  that was distributed as a part of the software suite {\em
  HEASOFT}.\footnote{see
  http://heasarc.gsfc.nasa.gov/docs/software/lheasoft} Briefly, for
  each observation, we simulated background events with the
  appropriate background model. The data were then filtered in the
  usual manner.\footnote{see the online \RXTE\ Cook Book at
  http://heasarc.gsfc.nasa.gov/docs/xte/recipes/cook\_book.html},
  which resulted in a list of Good Time Intervals (GTIs). Using the
  GTIs, we proceeded to extract a light curve from the data and
  background files, respectively. The background-subtracted light
  curve was rebinned to 1/4 sec for cross-correlation analysis.  The
  only X-ray data used in this work are those with an overlap with
  optical data.

\section{Orbital lightcurves}

\begin{figure}
\includegraphics[angle=90,width=3.5in]{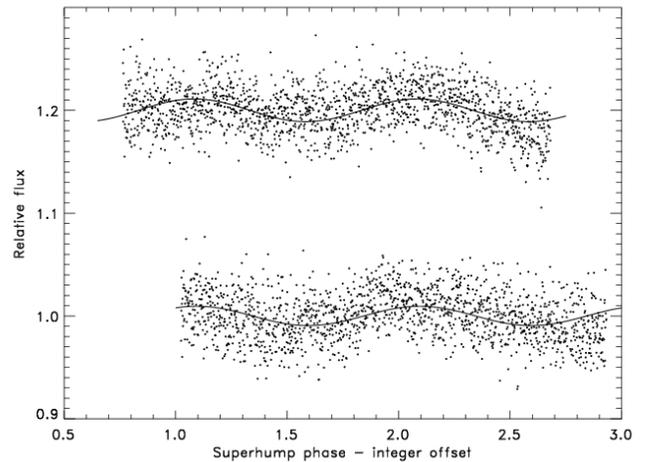}
\caption{Modulation in SARA photometry folded on preferred
1.0025\,$P_{\rm orb}$ period, likely due to a superhump.}
\label{SuperhumpFig}
\end{figure}

The best data for searching for an orbital or superhump modulation is
the SARA photometry as this has the longest periods of coverage, and
longer exposures suffer less scatter due to the large amplitude
flickering present.  A modulation on a period of about 4 hours is seen
on both nights at about the same full-amplitude of $\sim2.0$\,percent
(Fig.~\ref{SuperhumpFig}).  A period search including both nights of
data finds several closely spaced aliases.  One of them is at
0.17036(24)\,days or 1.0025(14)\,P$_{\rm orb}$ (assuming the period of
\citealt{Zurita:2002a}).  This is consistent with previously reported
superhump periods, e.g.\ 0.170529(6)\,days \citep{Uemura:2002a} but
does not securely rule out an orbital modulation.  Aliases at
1.09\,$P_{\rm orb}$ and 1.20\,$P_{\rm orb}$ are statistically slightly
preferred to that at the superhump period.  The similarity to previous
observations, points to the solution closest to the orbital period
being the correct one, however, as this is consistent with the finding
of \citet{Chou:2005a} of a very weak (0.02\,mag) modulation close to
the orbital (or superhump) period after January 18.  In contrast,
the superhump identified by \citet{Uemura:2002a}
throughout the 2000 outburst with exhibited a persistent modulation
with an average full-amplitude of about
7\,\%, so our observations and those of \citet{Chou:2005a} indicate a
weaker superhump in this outburst.

Where we have observations of at least one binary orbit in other data,
we examined the lightcurves for evidence of a similar superhump
modulation.  These other lightcurves are less well suited as rapid
variability is not averaged out.  We find no evidence of any such
modulations in either the optical or IR data.  The optical lightcurve
from January 16 places a $2\,\sigma$ (95\,\%) upper limit on the full
amplitude of a sinusoidal modulation at the orbital period (for any
phasing) of 3.5\,\%\ of the flux.  We attempted to analyze the IR
lightcurves in the same way and, while no consistent modulation was
apparent, the formal limits derived were much weaker.  This is a
consequence of the larger intrinsic variance due to flaring, and the
much lower duty cycle reducing the ability to average over many
flares.  Upper limits obtained were 7--16\,\%\ in $J$, 12--19\,\%\ in
$H$, and 17--24\,\%\ in $K_{\rm s}$.

\section{Power density spectra}
\label{PSDSection}

\begin{figure}
\includegraphics[angle=90,width=3.5in]{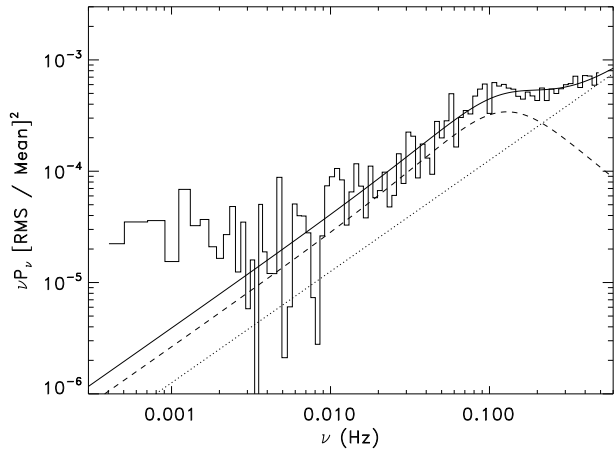}
\caption{Power density spectrum derived from fast photometry on 2005
  January 13, in unfiltered white light.  The dashed line is a
  single-Lorentzian component, the dotted line is the fitted
  white-noise level, and the solid line is the combined fit.  Fits
  only used data above 0.01\,Hz, where source variability is dominant.}
\label{ArgosFig}
\end{figure}

Our best time-sampling is in the fast optical photometry from 2005
January 13, 1\,s time-resolution with negligible deadtime.  In
Fig.~\ref{ArgosFig} we show the power-spectral density (PSD) of the
differential lightcurve between the target and comparison star.
Examining the PSDs of each star individually, the target exhibits
excess power above about 0.01\,Hz, even though it is brighter, while
below that both stars have the same power, likely due to transparency
variations and/or aperture losses.  We are therefore confident that
the power in the differential PSD above 0.01\,Hz should be due to real
variations in the target, apart from white-noise which dominates at
the very highest frequencies.  We have shown the PSD in the now common
$\nu P_{\nu}$ form and show a fit comprising a single Lorentzian
component plus white-noise \citep{Belloni:2002a}.  The Lorentzian
frequency parameters are a characteristic frequency $\nu_{\rm
max}=0.13$\,Hz and central frequency $\nu_0=0.065$\,Hz.  $\nu_{\rm
max}$ corresponds to the peak in $\nu P _{\nu}$ and is given by $\nu
_{\rm max} = \sqrt{\nu_0^2 + \Delta^2}$ where $\Delta$ is the
half-width at half maximum of the Lorentzian.  The difference from a
zero-centered Lorentzian is small, corresponding to a low coherence
$Q=0.11$.  The PSD derived is thus very similar to those seen during
the 2000 outburst, although the break frequency, $\nu_{\rm max}$, is
higher than seen then, when it evolved from about 0.03--0.08\,Hz
\citep{Hynes:2003a}.

If the PSD did not change substantially between 2005 January 13--19,
and is similar in the IR and optical, then this gives us information
about how well the IR observations resolve variability.  The single
fitted Lorentzian component has 90\,\%\ of its power on timescales
0.9--70\,s.  Thus most of the variability ($\ga90$\,\%) is on
timescales shorter than the $\sim50$\,s sampling time, and hence
consecutive points are effectively independent.  On the other hand,
most of the variability is on timescales longer than the IR exposure
time, hence the IR observations should not smooth out much of the
variability and should sample most of the dynamic range expected.

\section{X-ray optical correlations}

\begin{figure}
\includegraphics[angle=90,width=3.6in]{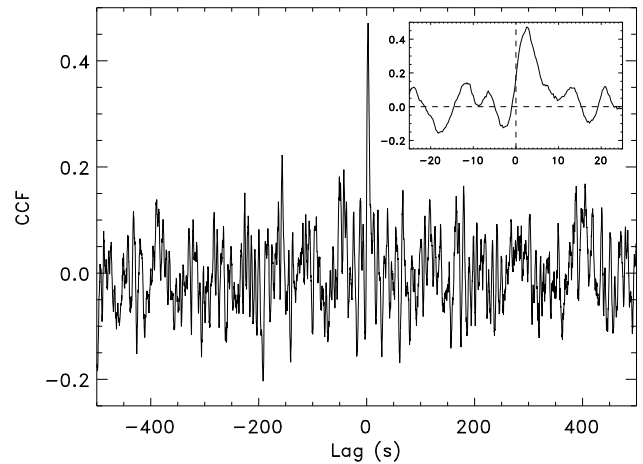}
\caption{Cross-correlation function between X-ray and 3\,s
time-resolution optical data from January 16.  The inset shows an
expanded view of the peak.}
\label{CCFFig}
\end{figure}

A small amount of overlapping X-ray and optical/IR data were obtained.
The poor duty cycle of the IR data resulted in only a handful of
points and no measurable correlation.  A correlation was seen
between X-ray and optical variability, based on a total of
approximately 11\,min of overlap with McDonald 2.7\,m White Guider
photometry.  Fig.~\ref{CCFFig} shows the cross-correlation function
(CCF) from 2005 January 16.  Since the optical exposure times were
quite long, this was calculated by averaging the X-ray data over the
duration of each optical exposure (after applying the lag).  The CCF
is not of high quality, and the 3\,s time-resolution (set by the
exposure duration, not the cycle time) limits the precision of the
information obtained, but a lagged correlation is clearly present.
The centroid and width of the CCF peak are comparable to those seen by
\citet{Kanbach:2001a} during the 2000 outburst although the lower
time-resolution of our data do not permit a detailed comparison.

\section{The spectral energy distribution (SED) of the IR noise}

\subsection{Observational results}

\begin{figure*}
\includegraphics[angle=90,width=6.4in]{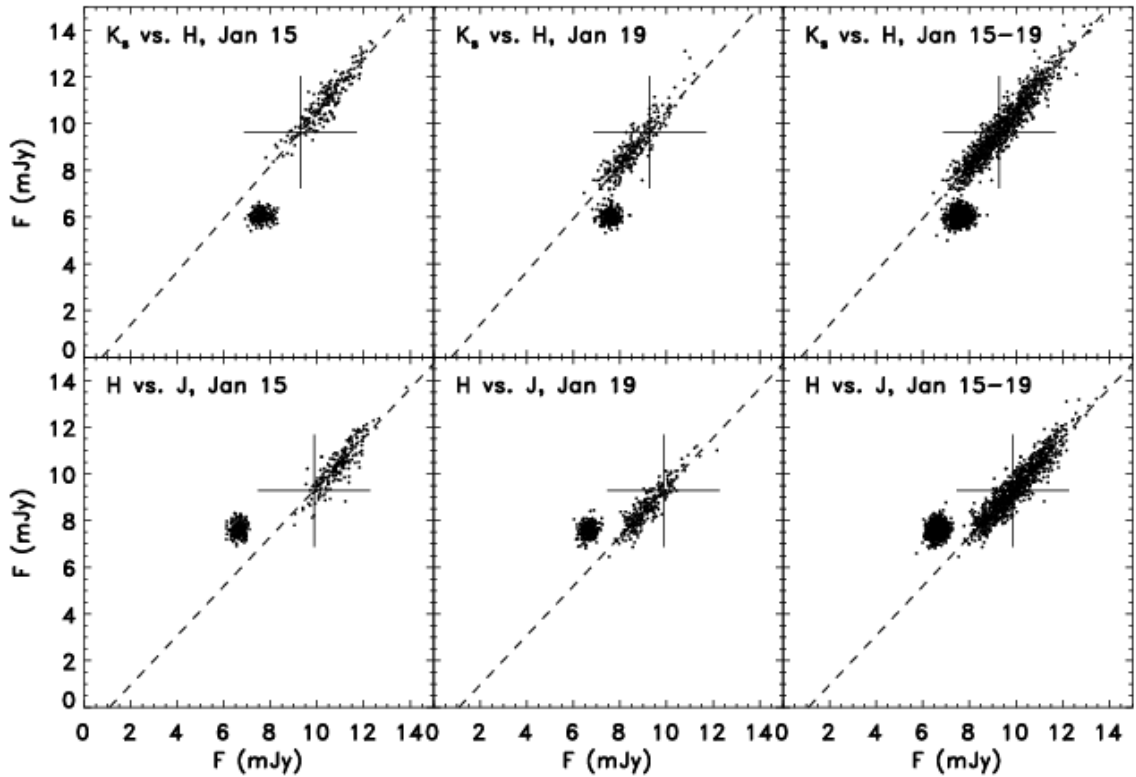}
\caption{$J$ vs.\ $H$ and $H$ vs.\ $K_{\rm s}$ fluxes for the first
night, the last night, and all nights with 2\,s data combined.  Two
clusters of points are present on each plot.  The diagonally dispersed
ones are \target, the circular clump are the comparison star.  The
dashed lines all show the same linear fit to the combined dataset (for
that color), to provide a point of reference in comparing the first
and last night.  The cross indicates the run averaged fluxes as a
point of reference.  The joint fit shown here is also only intended as
a reference point, and all analysis used night-by-night fits.}
\label{CorrelationFig}
\end{figure*}

Since $J$, $H$, and $K_{\rm s}$ are observed simultaneously, each
image-set provides an instantaneous IR SED.  Comparison of them allows
us to study the SED of the IR variability, even though we do not
resolve individual flares in time.  As a relatively large number of
such image-sets was obtained, we applied quite cautious standards in
filtering the data.  We excluded all sets of images in which i) the
seeing was greater than 1.7\,arcsec, ii) the sky subtraction in any
band was poor iii) the $J$ sky was high in twilight, iv) conditions
did not appear photometric (as measured by the comparison star), or v)
the target coincided with either a hot pixel or an $\alpha$-particle
hit on the detector.  Table~\ref{IRLogTable} lists the number of good
image-sets remaining after applying these cuts to the sample.  In
Fig.~\ref{CorrelationFig} we show examples of $J-H$ and $H-K_{\rm s}$
flux relations for the first and last nights and all nights with 2\,s
data combined.  The cuts applied do appear to have cleanly removed bad
data, leaving a well-defined correlations.  We also show equivalent
points for the comparison star.  For the latter each night's fluxes
have been divided by the median, so this is a measure of the scatter
in fluxes within a night.  If systematic effects, for example variable
extinction due to cirrus, were affecting our data then we would expect
the comparison star to exhibit correlated errors in different bands.
No such correlation is seen, suggesting the uncertainties are
dominated by random statistical noise.  The uncertainties in the
comparison (which should be larger than those in the brighter target)
are clearly much smaller than the observed variations in the target
fluxes.  The scatter about the correlation line in the target may be
statistical, however.

We performed the same analyses on the data from the third night, for
which 1\,s exposures were taken, to test if shorter exposures sampled
a larger dynamic range of the variability.  This does not obviously
appear to be the case, which is consistent with most of the power
being at lower frequencies (Section~\ref{PSDSection}).  Since these
data are noticably noisier than the other nights, we have not shown
them in the plots.  The overall distribution of fluxes is intermediate
between the preceding and following nights, with no obvious
differences other than the higher noise level.

Fig.~\ref{CorrelationFig} shows clear and repeatable correlations
between bandpasses.  Not only do the fluxes within a given night trace
out a well defined relationship, but successive nights appear to
follow approximately the same correlation, with the overall brightness
declining along the same line as the variations within a night.  This
points to a very consistent IR behavior and SED, and probably a single
spectral component dominant.  The consistency coupled with the
linearity over a factor of two in flux suggest that the shape of the
IR SED is not changing, only the overall flux level.

We can use the correlations to estimate the shape of the variable
component of the IR SED, since the slope of the correlation, for
example $dF_K / dF_H$, is a measure of the slope of the SED variations
between $H$ and $K$.  This measure is independent of any non-varying
zero-point, making this potentially a more sensitive discriminant
between models than the overall SED which may be a sum of several
spectral components (e.g.\ \citealt{Markoff:2001a};
\citealt{Yuan:2005a}).

\subsection{Power law models}
\label{PLSection}

The simplest model for the IR SED is a power-law.  We also tested
broken power-laws (i.e.\ allowing a different slope for $J-H$ and
$H-K_{\rm s}$), but did not find a convincing pattern (see
Table~\ref{PLFitTable}).  There is a tendency for the $J-H$ slope to
become flatter, and the $H-K_{\rm s}$ one to begin steeper
(corresponding to a concave spectrum), but suspiciously the $J-K_{\rm
s}$ slope stays approximately constant, and tracks a single power-law
fit to the full $JHK_{\rm s}$ datasets.  This suggests either that the
$H$ data are less reliable, or that adjacent bands simply do not
provide adequate leverage in wavelength.  We therefore see no
compelling reason to believe the spectral slope really changes across
the near-IR and proceed to fit a single power-law to the combined
$JHK_{\rm s}$ data for each night; the following discussion relates to
this joint fit.  Note that given three points in
wavelength, such a broken power-law model should be sufficient to
completely describe the wavelength dependence, and this test can also
be thought of as testing whether there is a significant and repeatable
curvature of the spectrum.

Free parameters in the unbroken power-law model are the assumed
zero-point fluxes in each band (we adopt the mean), the normalization
of the power-law for each image set, and the power-law slope.  For $n$
$JHK$ image sets there are thus $n+4$ free parameters and $2n-4$
degrees of freedom.  The resulting fits typically yielded a $\chi^2$
value of about 2 per degree of freedom, assuming only statistical
errors.  It is likely that the errors are underestimated in this way,
as we did not perform differential photometry and hence aperture
losses and small transparency variations could contribute.  To
approximately correct for this, we adopt the expedient of rescaling
the uncertainties (by a factor about $\sqrt 2$) to yield a minimum
$\chi^2$ of 1 per degree of freedom.  We then estimate the
single-parameter uncertainty in the slope, $\alpha$ (defined such that
$F_{\nu} \propto \nu^{\alpha}$), using the region encompassed by
$\Delta\chi^2 = 1$.  Our derived slopes are given in
Table~\ref{PLFitTable} and we also derive a weighted mean of the
slopes of $\alpha=-0.78$.  The four nights are consistent with their
weighted mean to within the quoted errors, and there is no overall
trend to the values, suggesting a constant spectral index over the
period observed.  

\begin{table}
\begin{tabular}{lccccc} 
\hline
Date   & $\alpha_{JH}$ & $\alpha_{HK}$ & $\alpha_{JK}$ &
$\alpha_{JHK}$ & $\chi^2 / {\rm dof}$\tablenotemark{a} \\
\hline
Jan 15 & $-0.75\pm0.12$ & $-0.73\pm0.11$ & $-0.78\pm0.07$ & $-0.77\pm0.06$ & 1.890 \\
Jan 16 & $-0.90\pm0.08$ & $-0.78\pm0.09$ & $-0.85\pm0.04$ & $-0.83\pm0.04$ & 1.822 \\
Jan 17 & $-0.30\pm0.16$ & $-1.04\pm0.15$ & $-0.67\pm0.08$ & $-0.70\pm0.07$ & 1.930 \\
Jan 18 & $-0.50\pm0.09$ & $-0.98\pm0.08$ & $-0.74\pm0.04$ & $-0.76\pm0.04$ & 1.996 \\
Jan 19 & $-0.48\pm0.07$ & $-1.02\pm0.08$ & $-0.74\pm0.06$ & $-0.74\pm0.05$ & 1.789 \\
\hline
\end{tabular}
\tablenotetext{a}{$\chi^2$ is for the joint $JHK$ fit.}
\caption{Power law fit parameters}
\label{PLFitTable}
\end{table}

Given that the statistical uncertainty is small, and the
night-to-night consistency is good, the dominant error will arise from
the systematic uncertainty in absolute calibration.  Using our
estimate of the absolute calibration uncertainties from
Section~\ref{IRObsSection}, we estimate using a Monte Carlo
calculation that the random uncertainty in the derived power-law index
will be about $\pm0.07$.  This assumes errors in $J$, $H$, and $K_{\rm
s}$ are independent.  If they are positively correlated then the
uncertainty in the power-law index will be reduced.  Only if there is
an anti-correlation between $J$ and $K_{\rm s}$ errors will we have
underestimated the uncertainty in the power-law index, and this is
unlikely.  This calibration uncertainty dominates over the other terms
so far discussed.  The final problem that introduces a systematic bias
rather than a random scatter is that the variability is redder than a
stellar SED.  We have treated the photometry as yielding monochromatic
fluxes at the bandpass center, but the measurements actually represent
weighted averages over the bandpass.  In practice this effect is not
large, however, so we neglect it.  \citet{Glass:1999a} estimate that
the corrections to near-IR fluxes to convert them to the monochromatic
flux at the bandpass center are only a few percent for power-laws of
spectral index between $-2$ and $+2$.  We estimate with some
simplistic simulations that for our $JHK_{\rm s}$ photometry, with the
spectral slope derived, the resulting error in the spectral slope is
likely to be less than 0.01, and much less than other uncertainties.
We have therefore not attempted a more rigorous treatment. 

In summary, all of our observations are consistent with a single
power-law of constant slope $\alpha=-0.78\pm0.07$, where the dominant
uncertainty is in the absolute calibration of the photometry.  

\subsection{Black body models}

A common paradigm for optical variability in X-ray binaries is that it
arises in reprocessed X-rays.  X-ray irradiation deposits energy at a
modest optical depth in the atmosphere of the disk or companion star.
This is then thermalized and emerges as optical and ultraviolet flux
which would, simplistically, be expected to have something close to a
black-body SED.  This interpretation was rejected based on detailed
analysis of the variability during the 2000 outburst
(\citealt{Kanbach:2001a}; \citealt{Hynes:2003a}).  Our results further
support this, as the variability we see is clearly too red to arise in
black body emission from the binary.  Black body fits analagous to
those discussed the preceding section yield temperatures
$\sim1500$\,K, too low for this to be a plausible interpretation.

\subsection{Optically thin thermal models}

\newcommand{\gta}{{\small\raisebox{-0.6ex}{$\,\stackrel
{\raisebox{-.2ex}{$\textstyle >$}}{\sim}\,$}}}

Although an optically thick thermal model is not plausible given the
redness of the variability, optically thin thermal emission could
still explain the spectral shape as the free-free emissivity increases
with wavelength, resulting in a redder SED than that of a black body.
  
\citet{Pearson:2005a} derived analytic expressions for the
time-dependent continuum spectra exhibited during flickering and
flaring events. The additional flux was modeled as arising from a
region of gas, with uniform temperature and Gaussian density profile,
expanding with a radial velocity proportional to the distance from the
centre.  Both free-free and bound-free emission mechanisms were
considered and the results of fitting to observations consistently
showed that the expansion exhibited isothermal evolution.  In the
optically thin limit, the observed flux is given by

\begin{eqnarray}
F & = & \frac{\pi a^{2} B}{2 d^{2}} \tau_{0} \label{eqn:fluxthin} \\
  & = &  \frac{\pi a^{2} B}{2 d^{2}} 
\left(
\frac{\kappa_{1} M^{2}}{\sqrt{2\pi^{5}} a^{5} T^{\frac{1}{2}}}
\frac{1-e^{-\frac{h\nu}{k T}}}{\nu^{3}}
\right) \label{eqn:tau0sub}\\
  & = & e^{-\frac{h\nu}{k T}}
\left(
\frac{h \kappa_{1} M^{2}}{\sqrt{2 \pi^{3}} c^{2} d^{2} a^{3} T^{\frac{1}{2}}}
\right) \label{eqn:longform}\\
  & \equiv & \alpha_{\nu} f
\end{eqnarray} 
where $a$ is the lengthscale of the Gaussian ($=\sqrt{2}$s.d.), $B$ is
the Planck function, $\tau_{0}$ is the optical depth through the
centre of the expansion, $d$ is the distance to the object, $T$ is the
temperature of the region and $M$ its mass. The term $\kappa_1$ is a
constant that depends on the composition and the emission mechanism
(i.e.\ whether free-free or bound-free).  $\alpha_{\nu}$ and $f$
encapsulate a number of terms from (\ref{eqn:longform}), while
separating the wavelength dependence into $\alpha_{\nu}$ alone.

Since the quantities contributing to $f$ are fixed for a given
observation of a particular flicker, the ratio of flux in each
waveband is given by the ratio of the values for the
$\alpha_{\nu}$. Experience suggests that not only is $T$ constant for
a particular event but is also fairly consistent between flickers. The
linear relationship between fluxes in the different wavebands can thus
be understood as a reflection of the linear relationship between the
flux and the parameter $f$.

For each data triple, consisting of the flux in the $JHK_{\rm s}$
wavebands, we can derive a best estimate instantaneous value for
$f$. We can then fit a straight line to $F(f)$ in each waveband
simultaneously and extract a best fitting value for $\alpha_{J}$ (or
equivalently $\alpha_{H}$ or $\alpha_{K}$) which in turn is a simple
function of $T$. In practice, rather than a single parameter, we also
have to allow for a zero point offset in the flux in each waveband to
account for constant or slowly varying contributions from other parts
of the sytem.  The derived values of $T$ are given in
table~\ref{tab:tfits} and have $\chi^2/\nu$ similar to the power-law
fits in Section~\ref{PLSection}.

\begin{table}
\begin{center}
\begin{tabular}{ccc} \hline
Date & Temperature
& $\chi^{2}/{\nu}$ 
\\
     &     (K)     & \\
\hline
Jan 15 &  11~500    & 1.888 \\
Jan 16 &  10~600    & 1.819 \\
Jan 17 &  12~800    & 1.940 \\
Jan 18 &  11~700    & 2.008 \\
Jan 19 &  12~000    & 1.797 \\
\end{tabular}
\end{center}
\caption{Derived temperature from the flickering occuring on each
  night for an optically thin thermal model.}
\label{tab:tfits}
\end{table}

As a consistency check, we can insert the condition that material be
optically thin ($\tau_0<1$) into (\ref{eqn:fluxthin}). This gives us,
\begin{eqnarray}
a_{\rm thin} & > & \sqrt{\frac{2 d^2 F}{\pi B}} \\
             & \gta & 3\times10^{9} \mbox{~~m}   
\end{eqnarray}
using $d=1.8$~kpc, $T\sim12~000$~K and $F_{K}\sim20$~mJy.
Unfortunately then, while the thermal models do manage to reproduce
the data equally as well as a power law, they require an emiting
region comparable in size to the entire binary
($a\sim2\times10^9$\,m).  It is hard to envisage how this could arise
from material within the primary's Roche Lobe.

This difficulty suggests that, despite the attractiveness of the fit,
the emission mechanism, in this case, is not thermal in origin, and
hence that our original power-law fits are probably most meaningful.

\section{Discussion}

The properties of the variability we observe are very similar to those
seen during the 2000 outburst, for example the prompt correlation with
X-rays, the shape of the PSD, and the increasing variability at longer
wavelengths.  Both \citet{Kanbach:2001a} and \citet{Hynes:2003a}
concluded that synchrotron emission was the most likely origin of the
variability.  Our multicolor observations provide strong support for
this interpretation.  We have argued that a thermal model, whether
optically thick or thin, cannot adequately explain the combination of
spectral shape, luminosity, and variability timescales observed.  The
inferred IR spectral slope, $\alpha \simeq -0.78$ is, on the other
hand, quite appropriate for optically thin synchrotron emission.

In the context of the jet model for the broad-band SED of \target\
presented by \citet{Markoff:2001a}, this implies that the IR
variability, and probably also the optical and some UV, originate from
the ``optically thin post-shock jet'' component.  This may indicate a
difference from the models of \citet{Markoff:2001a} in which the
near-IR exhibits a flat spectrum, as this implies that the break
between optically thin and partially self-absorbed synchrotron occurs
at longer wavelengths than the near-IR.  The slope we derive, however,
$\alpha \simeq -0.78$, appears in good agreement with these models.
Our observations therefore provide new tests of these models, allowing
isolation of the optically thin jet emission from contamination by the
disk.

\citet{Yuan:2005a} consider a coupled accretion-jet model. In this
model, the IR emission is mostly due to radiation from the jets,
although the contribution from hot accretion flows might not be
negligible. The jet contribution is associated with synchrotron
emission from electrons that are accelerated by internal shocks in the
jets. The spectral distribution of the electrons is assumed to be of
power-law shape, $N(E) \propto E^{-p}$ for $E_{\rm min} \le E \le
E_{\rm max}$, where $p=2.24$. For optically thin synchrotron emission
one then expects that the spectrum of the radiation is also of
power-law shape, for the most part, with a spectral slope
$\alpha=(1-p)/2=-0.62$, which is already rather close to that which we
observe. Moreover, for \target, the IR emission may be associated with
electrons near $E_{\rm max}$, so its spectrum deviates from the power
law and is steeper \citep{Yuan:2005a}.

\section{Conclusions}

We have performed optical, IR, and X-ray observations of rapid
variability in \target\ during its 2005 outburst.  Many
characteristics are similar to the 2000 outburst, although there are a
few key differences.  Superhumps, if present, are weaker than in 2000.
The variability also seems concentrated to higher frequencies.  Our
major novel result is simultaneous $J,H,K_{\rm s}$ photometry allowing
us to isolate the SED of the IR variability.  We find this is red, and
can be well fitted by a power-law, $F_{\nu} \propto \nu^{\alpha}$,
where $\alpha=-0.78\pm0.07$.  This result is consistent with optically
thin synchrotron emission but hard to explain with thermal emission.
We consider this to be strong evidence in favor of the interpretation
of the variability as arising in synchrotron emission, most likely
from a jet.  Unlike attempts to model the average spectral energy
distribution (e.g.\ \citealt{Markoff:2001a}; \citealt{Yuan:2005a}),
the variability isolates the synchrotron emission from the disk
emission allowing us to measure of the slope of the optically thin
component directly and show that the spectral break to self-absorbed
emission must occur at longer wavelengths than the near-IR, at least
at the time our observations were made.

\acknowledgments

We are grateful to Michael Merrill for technical assistance with SQIID
and John Robertson for assistance with the SARA
observations.  W.C.\ acknowledges support from NASA grant NNG04GI54G.
This publication makes use of data products from the Two Micron All
Sky Survey, which is a joint project of the University of
Massachusetts and the Infrared Processing and Analysis
Center/California Institute of Technology, funded by the National
Aeronautics and Space Administration and the National Science
Foundation.  This work has made use of the NASA Astrophysics Data
System Abstract Service.

\clearpage

%% Use the figure environment and \plotone or \plottwo to include 
%% figures and captions in your electronic submission.

%\begin{figure}
%\plotone{f1.eps}
%\caption{This is the first figure and it uses sgi9259.eps as
%its EPS figure file. \label{fig1}}
%\end{figure}

%\clearpage 

\end{document}